\documentclass[twocolumn,prb,aps,showpacs,amsmath,amssymb]{revtex4-1}

\usepackage{graphicx}
\usepackage{dcolumn}
\usepackage{bm}

\renewcommand{\vec}[1]{{\bf#1}}

\begin{document}

\title{Asymmetric Cherenkov acoustic reverse in topological insulators}

\author{Sergey Smirnov}
\affiliation{Institut f\"ur Theoretische Physik, Universit\"at Regensburg,
  D-93040 Regensburg, Germany}

\date{\today}

\begin{abstract}
A general phenomenon of the Cherenkov radiation known in optics or acoustics
of conventional materials is a formation of a forward cone of, respectively,
photons or phonons emitted by a particle accelerated above the speed of light
or sound in those materials. Here we suggest three-dimensional topological
insulators as a unique platform to fundamentally explore and practically
exploit the acoustic aspect of the Cherenkov effect. We demonstrate that
applying an in-plane magnetic field to a surface of a three-dimensional
topological insulator one may suppress the forward Cherenkov sound up to zero
at a critical magnetic field. Above the critical field the Cherenkov sound
acquires pure backward nature with the polar distribution differing from the
forward one generated below the critical field. Potential
applications of this asymmetric Cherenkov reverse are in design of low energy
electronic devices such as acoustic ratchets or, in general, in low power
design of electronic circuits with a magnetic field control of the direction
and magnitude of the Cherenkov dissipation.
\end{abstract}


\maketitle
\section{Introduction}\label{intro}
The Cherenkov radiation discovered experimentally by Cherenkov
\cite{Cherenkov} in optics of transparent media and theoretically
explained later by Tamm and Frank \cite{Tamm} represents a general and
important channel of energy dissipation. This kind of dissipation arises
whenever fast particles propagate in media with which they interact via
certain microscopic mechanisms. This dissipation mechanism is not restricted
to only optical media where particles exciting photons have velocities in
excess of the speed of light in those media. It also appears in solids where
particles move faster than sound and, as a result, excite lattice vibrations
or phonons. In both cases the photons or phonons are distributed within a
forward cone centered around the momentum of the particle producing the
Cherenkov light or sound.

Focusing on the Cherenkov sound, or the acoustic Cherenkov effect, one may
distinguish three aspects particularly important in practice. The first aspect
is related to energy losses in electronic devices. Indeed, in any solid
electrons are coupled to the lattice. The strength of this coupling is
temperature independent. Therefore, at any temperature fast electrons emit
phonons, {\it i.e.}, lose their energy via the Cherenkov dissipation limiting
in this way the efficiency of devices. The second aspect of the acoustic
Cherenkov effect is its use in building acoustic devices. Here implementations
include acoustic amplifiers based on Si/SiGe/Si heterostructures
\cite{Komirenko_2000}, the GaAs technology \cite{Zhao_2009}, piezoelectric
semiconductors \cite{Willatzen_2014} as well as terahertz sound sources based
on graphene \cite{Zhao_2013}. Finally, the third aspect of the Cherenkov sound
consists in investigation of the properties of the media where it propagates
because the character of the Cherenkov sound strongly depends on these
properties. This aspect of the Cherenkov sound has been, {\it e.g.}, exploited
in Ref. \onlinecite{Kovrizhin_2001} to study properties of ultracold Bose
gases.

In the above first two practical aspects of the Cherenkov acoustic effect it
is essential to have a, possibly, simple control over the magnitude and
direction of the generated sound. In particular, for the device efficiency it
is crucial to reduce the energy losses and it is thus attractive to get an
external access to the Cherenkov radiation with the possibility to completely
close this dissipation channel. Alternatively, an external control is
invaluable to direct the Cherenkov sound along specific directions and block
its propagation in some others. In this situation one meets a fundamental
problem of overcoming the forward Cherenkov cone which in an optical setup
has been addressed in Ref. \onlinecite{Joannopoulos} within photonic crystals.
In these systems a possibility to generate pure backward Cherenkov light has
been demonstrated at the expense of the system's spatial inhomogeneity. Also
an acoustic implementation of backward Cherenkov sound has been explored in
electronic systems with spin-orbit interactions which is particularly
appealing for future spintronic devices. For example, in two-dimensional
electron gases with the Bychkov-Rashba \cite{Bychkov} spin-orbit interaction
interchiral transitions are responsible for the generation of the sound
outside \cite{Smirnov_2011} the Cherenkov cone. The strong spin-orbit
interaction in topological insulators \cite{Hasan_2010,Qi_2011} and the
helical \cite{Wu_2006,Xu_2006} nature of their surface states offer an
alternative \cite{Smirnov_2013} possibility to generate backward Cherenkov
sound due to only intrachiral transitions between the anisotropic
\cite{Fu_2009} surface helical states.

The Cherenkov sound in these spin-orbit coupled systems represents fundamental
interest and can be exploited to study the properties of these
systems. However, its practical use in electronic devices could be limited
because of some drawbacks. First, although backward Cherenkov sound appears,
it is not pure because together there always appears forward Cherenkov
sound. Second, the magnitude of the Cherenkov sound is fixed by the system
parameters which are difficult or even impossible to switch during device
operation. Additionally, in the case of topological insulators the required
anisotropy of the helical surface states starts to play a role at rather large
energies and momenta. This could limit the use of the anisotropy in
applications of topological insulators in the future generation of integrated
electronic circuits operating at low energies.

To avoid the above mentioned drawbacks we propose in this paper a simple way
to externally control the Cherenkov dissipation by means of an in-plane
magnetic field. For this type of control surface helical states of
three-dimensional topological insulators turn out to be most suitable
especially at low energies relevant for applications of topological insulators
in low power design electronics. One of the relevant device applications we
discuss is an acoustic ratchet. This acoustic spin-orbit ratchet may be
considered as a phonon alternative to the electron spin-orbit ratchet proposed
in Ref. \onlinecite{Smirnov_2008} within a two-dimensional Bychkov-Rashba
electron gas.

The paper is organized as follows. In Section \ref{CR3DTI} we qualitatively
describe the picture of the Cherenkov acoustic reverse where the forward sound
reverses to the backward one when an in-plane magnetic field exceeds a
critical value. The actual polar distribution of the forward and backward
Cherenkov sound is found in Section \ref{2DCSD} where we demonstrate that the
Cherenkov reverse has an asymmetric character. We further explain the physical
reason of this asymmetry and suggest to use it in electronic devices such as
acoustic ratchets. In Section \ref{concl} we conclude the paper.

\section{Cherenkov reverse on a surface of a three-dimensional topological
  insulator}\label{CR3DTI}
The low energy surface helical states of a three-dimensional topological
insulator are described by the Dirac Hamiltonian,
\begin{equation}
\hat{H}_0=v(\hat{p}_x\hat{\sigma}_y-\hat{p}_y\hat{\sigma}_x),
\label{Dirac_Ham}
\end{equation}
where $v$ is the Dirac velocity and $\hat{p}_i$, $\hat{\sigma}_i$, $i=x,y$,
are the momentum and spin operators. The important difference of this Dirac
Hamiltonian from the one describing graphene \cite{Castro_Neto_2009} is that
in Eq. (\ref{Dirac_Ham}) the particle momentum is coupled to the real spin and
not to the pseudo-spin describing the lattice degree of freedom in graphene.

Exactly this circumstance provides an opportunity to control the Cherenkov
dissipation in a three-dimensional topological insulator applying an in-plane
magnetic field to its surface. Indeed, such a field is described by the Zeeman
contribution $\hat{H}_Z$ to the total Hamiltonian,
\begin{equation}
\begin{split}
&\hat{H}=\hat{H}_0+\hat{H}_Z,\\
&\hat{H}_Z=\frac{1}{2}g\mu_\text{B}\hat{\vec{\sigma}}\vec{H},
\end{split}
\label{Zeeman_Ham}
\end{equation}
where $g$ is the $g$-factor, $\mu_\text{B}$ is the Bohr magneton,
$\hat{\vec{\sigma}}=\vec{i}\hat{\sigma}_x+\vec{j}\hat{\sigma}_y$ and
$\vec{H}=\vec{i}H_x+\vec{j}H_y$. Therefore, the position of the Dirac point,
which is the minimum of the conduction band and the maximum of the valence
band, becomes magnetic field dependent as can be seen from the energy-momentum
dispersion relation,
\begin{equation}
\begin{split}
\epsilon_\mu(\vec{p})=
\mu\biggl[v^2\vec{p}^2&+v|\vec{p}|g\mu_\text{B}|\vec{H}|\sin(\phi_\vec{H}-\theta_\vec{p})+\\
&+\biggl(\frac{1}{2}g\mu_\text{B}|\vec{H}|\biggl)^2\biggl]^\frac{1}{2},
\end{split}
\label{Energy_momentum}
\end{equation}
where $\mu=\pm 1$ is the chiral index ($+1$-conduction band, $-1$-valence
band), $\phi_\vec{H}$ and $\theta_\vec{p}$ are, respectively, the angles of
the in-plane magnetic field and the helical particle momentum with respect to
the $x$-axis.
\begin{figure}
\includegraphics[width=8.0 cm]{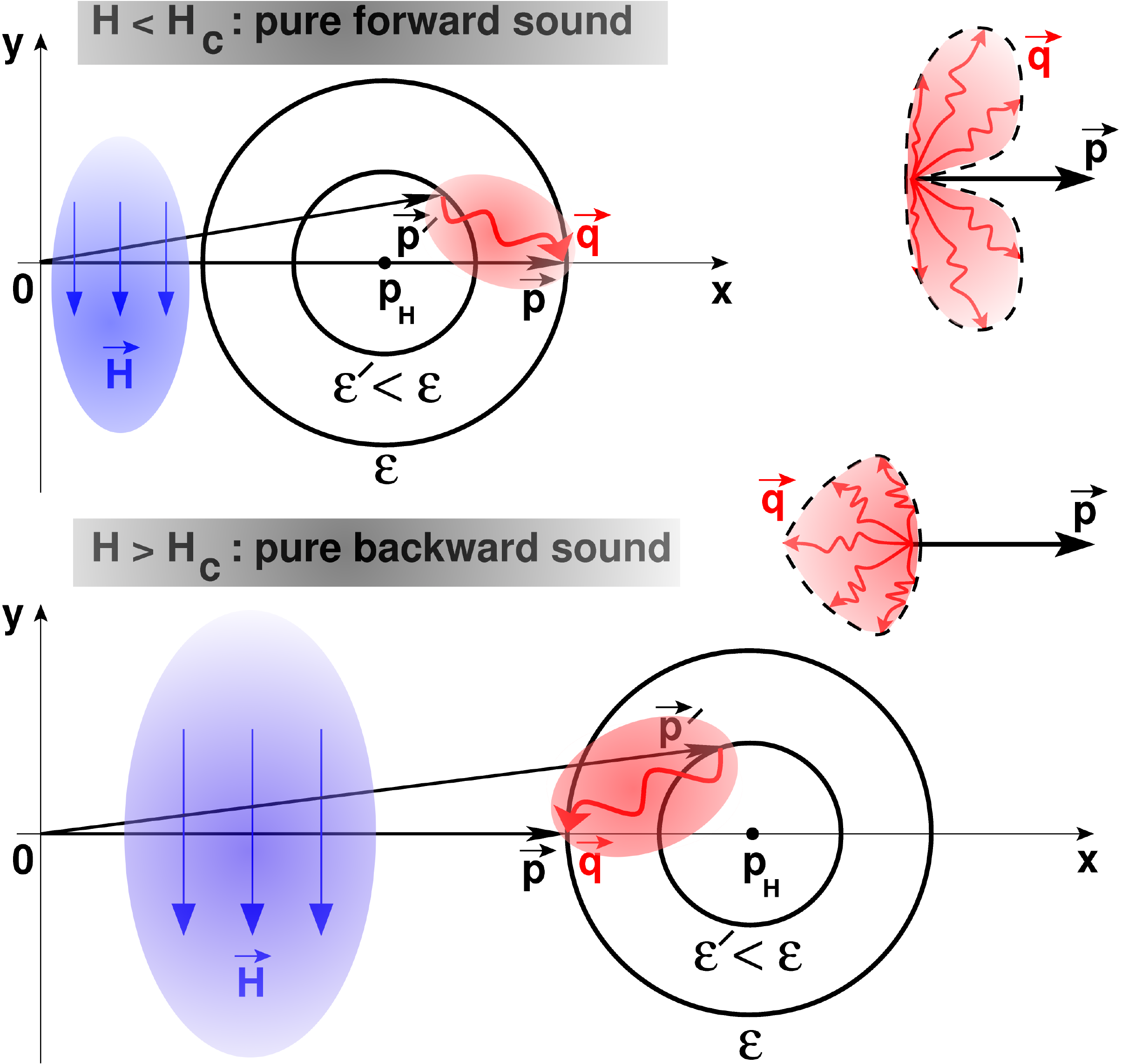}
\caption{\label{figure_1} The schematic picture of the Cherenkov acoustic
  reverse on a surface of a three-dimensional topological insulator subject to
  an in-plane magnetic field. The helical particle's momentum and energy
  before exciting the Cherenkov sound are $\vec{p}$ (black thick horizontal
  arrows) and $\varepsilon$ while after emitting a phonon with momentum
  $\vec{q}$ (red wavy arrows) they become $\vec{p}'$ (black thick inclined
  arrows) and $\varepsilon'$. An in-plane magnetic field (blue vertical
  arrows) is applied in the direction opposite to the $y$-axis. The black
  circles represent low energy isotropic surfaces of constant energy given by
  Eq. (\ref{Energy_momentum}). Their center $p_H$ (black dot) is the magnetic
  field dependent Dirac point. In the upper part $p_H<|\vec{p}|$ resulting in
  pure forward Cherenkov sound. In the lower part $p_H>|\vec{p}|$. In this
  case the Cherenkov sound reverses and acquires pure backward nature. In both
  cases the Cherenkov sound distribution is shown as the shaded red area with
  phonon momenta $\vec{q}$ (red wavy arrows) inside.}
\end{figure}

In topological insulators the Dirac velocity $v$ is usually much larger than
the sound velocity $c$. For example, for Bi$_2$Te$_3$ these velocities are
$v\approx 3.87\times 10^5$ m/s and $c\approx 2.84\times 10^3$ m/s. The
relation $v\gg c$ implies \cite{Smirnov_2013} that the Cherenkov sound may be
generated only due to transitions
$|\vec{p}\mu\rangle\rightarrow|\vec{p}'\mu'\rangle$ with $\mu=\mu'$.

These two remarkable features specific to topological insulators, the magnetic
field shift of the Dirac point and the intrachiral nature of the Cherenkov
sound, provide a simple possibility for an external control of the Cherenkov
dissipation.
\begin{figure}
\includegraphics[width=8.0 cm]{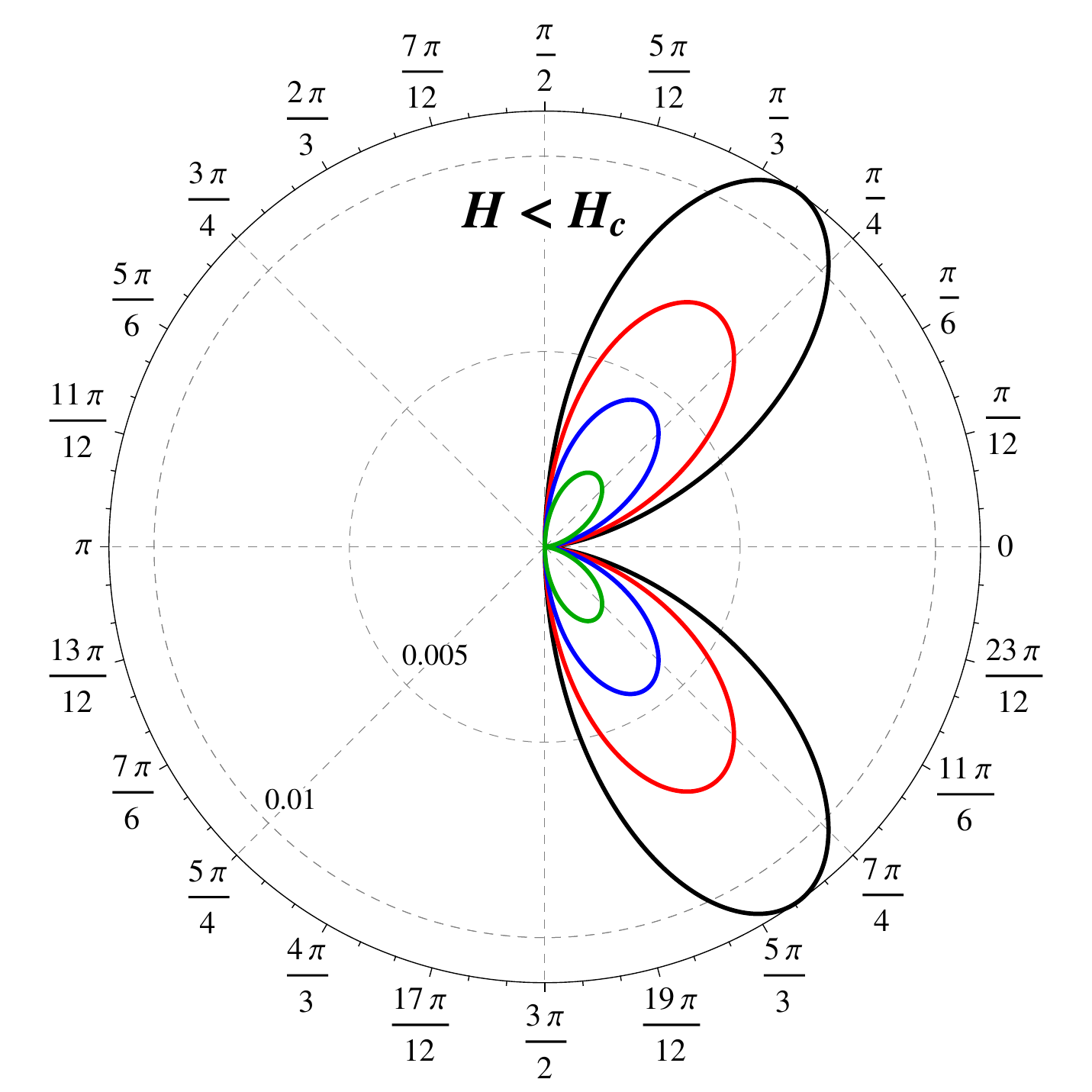}
\caption{\label{figure_2} The polar distributions of the Cherenkov sound
  intensity $W(\phi)$ on a surface of Bi$_2$Te$_3$ for the case
  $|\vec{H}|<H_c$. The magnitude of the helical particle momentum
  $|\vec{p}|=4.05\times10^{-27}$ is chosen to be small to stay within low
  energies well captured by the Dirac isotropic model. The polar distributions
  correspond to different magnitudes of the magnetic field, {\it i.e.}, to
  different values of the parameter $h$: $0.0$ (black), $50.0$ (red), $100.0$
  (blue), $150.0$ (green). The critical value of $h$ corresponding to $H_c$ is
  $h_c=272.0$. The Cherenkov sound is purely forward for all $h<h_c$. When $h$
  increases, $W(\phi)$ decreases and totally vanishes at $h=h_c$.}
\end{figure}
\begin{figure}
\includegraphics[width=8.0 cm]{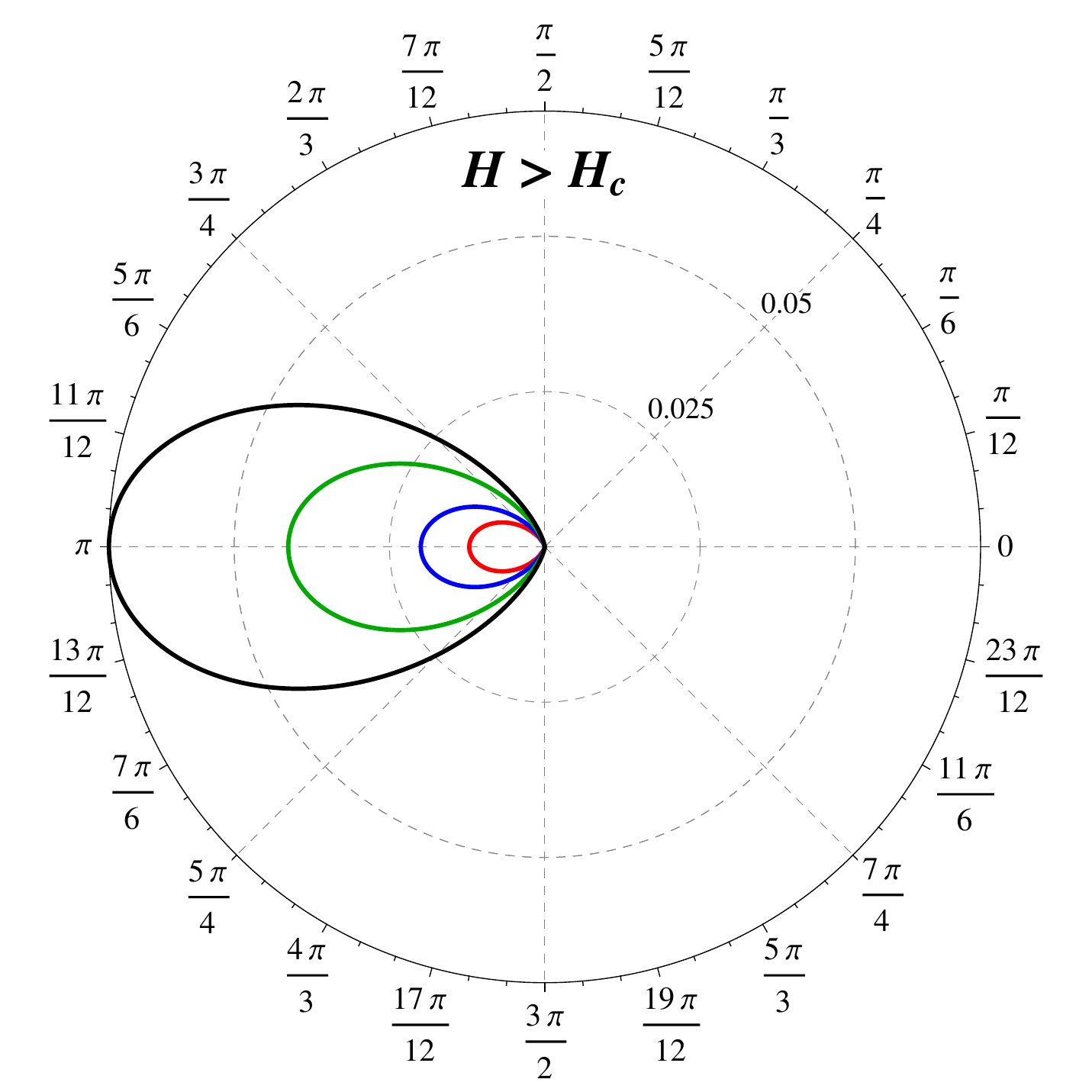}
\caption{\label{figure_3} The polar distributions of the Cherenkov sound
  intensity $W(\phi)$ on a surface of Bi$_2$Te$_3$ for the case
  $|\vec{H}|>H_c$. The helical particle momentum is the same as for
  Fig. \ref{figure_2}. The polar distributions correspond to $h=450.0$ (red),
  $500.0$ (blue), $600.0$ (green), $700.0$ (black). The Cherenkov sound is of
  pure backward nature for all $h>h_c$.}
\end{figure}

To qualitatively explain the idea let us consider the Cherenkov sound excited
by a helical particle in the conduction band. The particle has energy
$\varepsilon$ and momentum $\vec{p}$ along the $x$-axis
($\theta_\vec{p}=0$). An in-plane magnetic field is applied perpendicular to
the vector $\vec{p}$ ($\phi_\vec{H}=-\pi/2$). As demonstrated in
Fig. \ref{figure_1}, the low energy Dirac-like isotropic character of the
constant energy surfaces does not change in the presence of an in-plane
magnetic field which only shifts the Dirac point from zero to $p_H$. The Dirac
point is a special point of the helical particle energy-momentum dependence,
Eq. (\ref{Energy_momentum}), and tuning the magnitude of the magnetic field to
the critical value $|\vec{H}|=H_c$, such that $p_H=|\vec{p}|$, changes the
behavior of the system. Indeed, as one can see from Fig. \ref{figure_1}, in
the case $|\vec{H}|<H_c$ one has $p_H<|\vec{p}|$ and the momentum $\vec{p}$ of
the helical particle exciting the Cherenkov sound touches the corresponding
constant energy surface from the inside of the Dirac cone. In such a situation
the energy and momentum conservation admits only such helical particle's final
states, characterized by energy $\varepsilon'$ and momentum $\vec{p}'$, which
result in purely forward Cherenkov sound located inside the Cherenkov cone. On
the other side, when $|\vec{H}|>H_c$, the Dirac point satisfies
$p_H>|\vec{p}|$ and one is in the situation where $\vec{p}$ touches the
constant energy surface from the outside of the Dirac cone. As shown in
\begin{figure}
\includegraphics[width=8.0 cm]{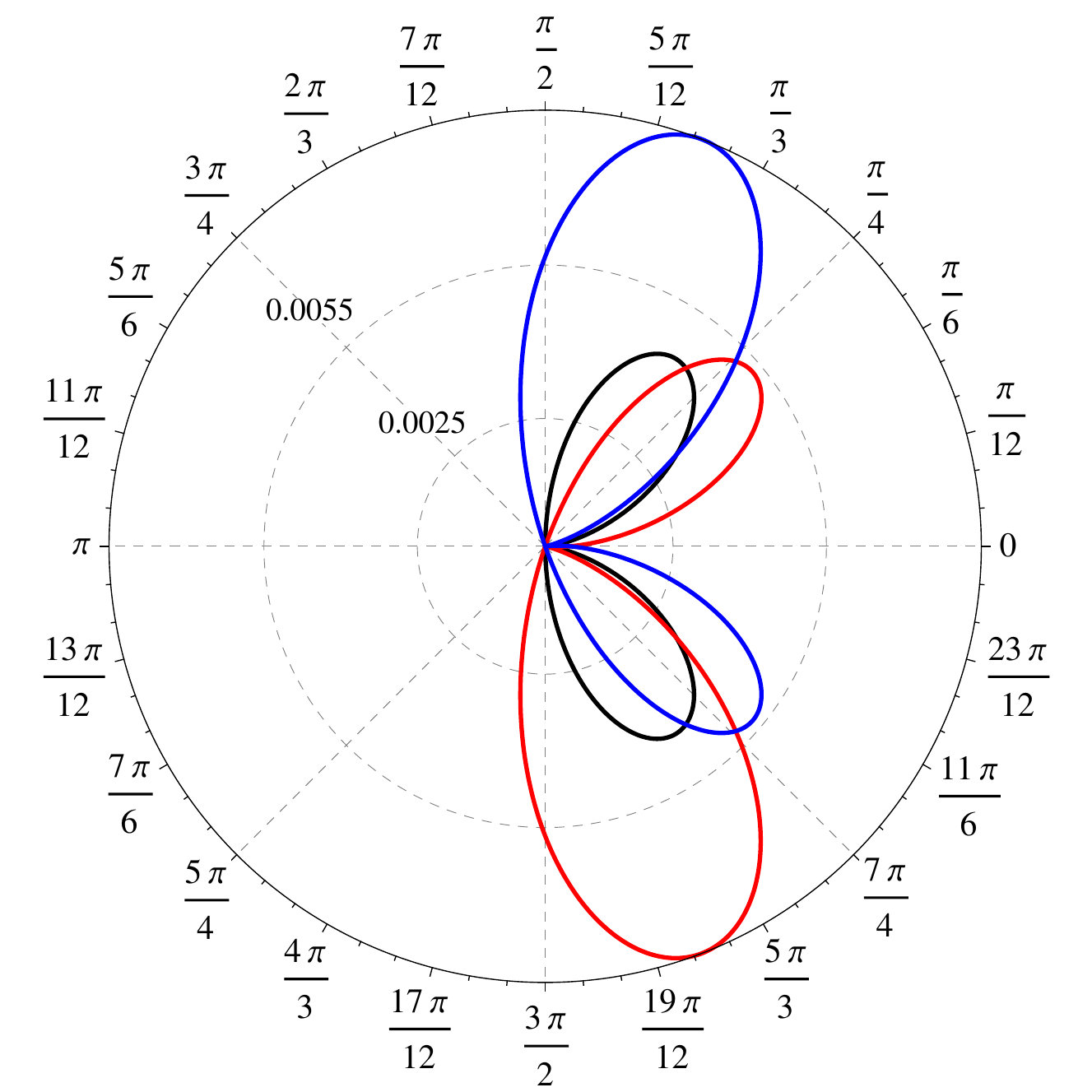}
\caption{\label{figure_4} The polar distributions of the Cherenkov sound
  intensity $W(\phi)$ on a surface of Bi$_2$Te$_3$ for the case
  $|\vec{H}|<H_c$. The helical particle momentum is the same as for
  Fig. \ref{figure_2}. The polar distributions correspond to $h=100.0$ and
  $\phi_\vec{H}=-\pi/2$ (black), $\phi_\vec{H}=-\pi/4$ (red) and $\phi_\vec{H}=-3\pi/4$ (blue).
  The Cherenkov sound is mainly forward but narrow backward angular sectors
  appear for $\phi_\vec{H}=-\pi/4$ and $\phi_\vec{H}=-3\pi/4$.}
\end{figure}
Fig. \ref{figure_1}, under such circumstances the energy and momentum
conservation reverts the Cherenkov sound placing it outside the Cherenkov
cone, {\it i.e.}, enforces it to acquire pure backward nature.

\section{Two-dimensional distribution of the forward and backward Cherenkov sound}\label{2DCSD}
To quantitatively verify this picture and to find out the actual distribution
of the Cherenkov sound and its dependence on the in-plane magnetic field we
calculate the imaginary part of the self-energy describing the interaction
\cite{AGD} between helical particles and phonons,
\begin{equation}
\begin{split}
&\hat{H}_\text{ph}=\sum_{\vec{k}}\hbar\omega(\vec{k})(b^\dagger_\vec{k}b_\vec{k}+1/2),\\
&\hat{H}_\text{el-ph}=g_{ph}\sum_\sigma\int d\vec{r}\hat{\psi}^\dagger_\sigma(\vec{r})\hat{\psi}_\sigma(\vec{r})\hat{\varphi}(\vec{r}),\\
&\hat{\varphi}(\vec{r})=i\sum_{\vec{k}}\sqrt{\frac{\hbar\omega(\vec{k})}{2V}}\biggl[\exp\biggl(i\frac{\vec{k}\vec{r}}{\hbar}\biggl)b_\vec{k}-h.c.].
\end{split}
\label{Ham_eph}
\end{equation}
In Eq. (\ref{Ham_eph}) $b^\dagger_\vec{k}$, $b_\vec{k}$ are the phonon
creation and annihilation operators, respectively, the phonon spectrum is
$\hbar\omega(\vec{k})=c|\vec{k}|$, $g_{ph}$ is the strength of the coupling to
phonons, $V$ is the volume and $\hat{\psi}^\dagger_\sigma(\vec{r})$,
$\hat{\psi}_\sigma(\vec{r})$ are, respectively, the helical particle creation
and annihilation field operators.
\begin{figure}
\includegraphics[width=8.0 cm]{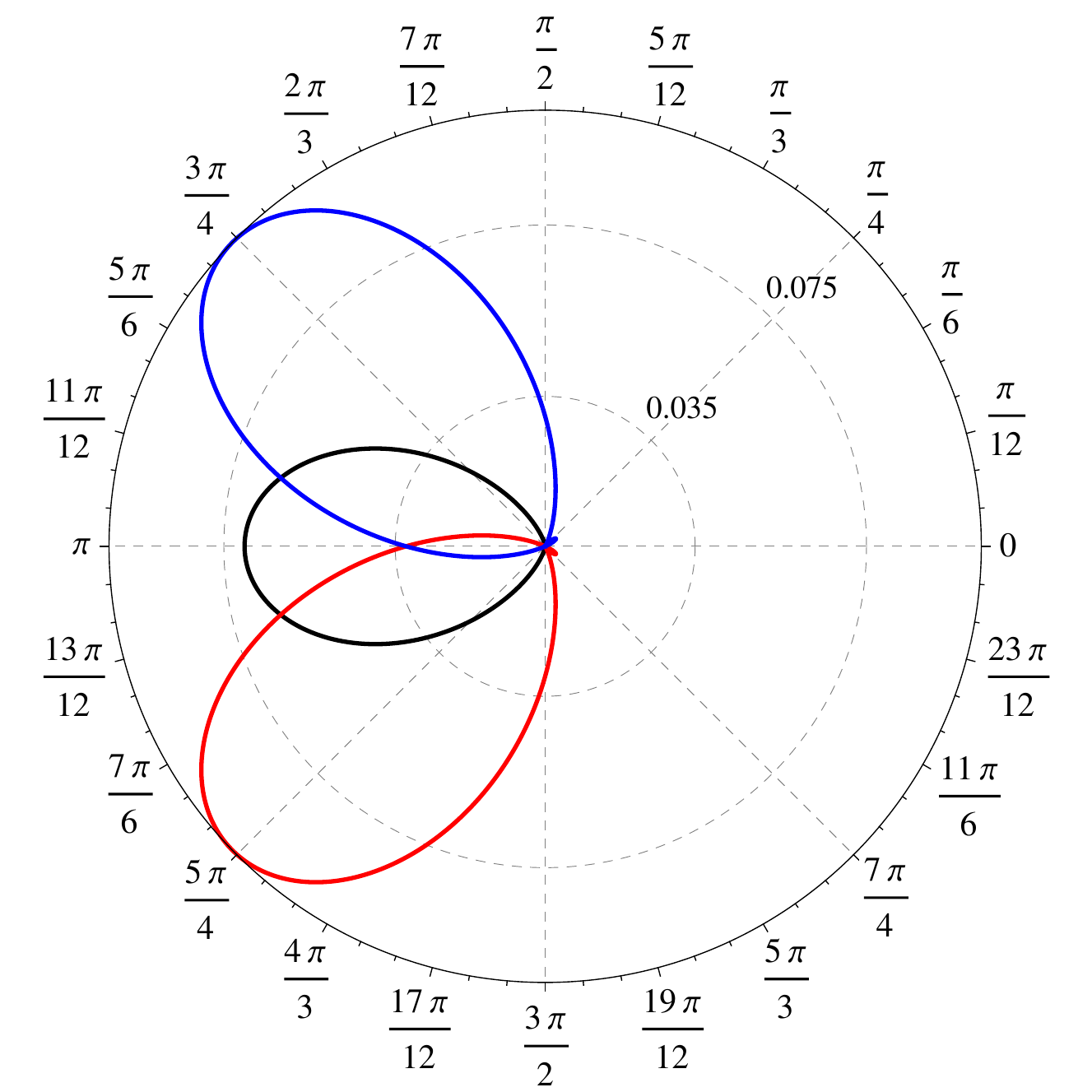}
\caption{\label{figure_5} The polar distributions of the Cherenkov sound
  intensity $W(\phi)$ on a surface of Bi$_2$Te$_3$ for the case
  $|\vec{H}|>H_c$. The helical particle momentum is the same as for
  Fig. \ref{figure_2}. The polar distributions correspond to $h=700.0$ and
  $\phi_\vec{H}=-\pi/2$ (black), $\phi_\vec{H}=-\pi/4$ (red) and $\phi_\vec{H}=-3\pi/4$ (blue).
  The Cherenkov sound is mainly backward but narrow forward angular sectors
  appear for $\phi_\vec{H}=-\pi/4$ and $\phi_\vec{H}=-3\pi/4$.}
\end{figure}

In the second order in $g_{ph}$ a derivation similar to the ones in
Refs. \onlinecite{Smirnov_2011,Smirnov_2013,LS} allows one to represent the
imaginary part of the self-energy $\Sigma(\epsilon_{\vec{p}+})$ in the form
\begin{equation}
\text{Im}[\Sigma(\epsilon_{\vec{p}+})]=-\frac{g_{ph}^2\vec{p}^2}{8\pi\hbar^3}\int_{-\pi}^\pi
d\phi W(\phi),
\label{Im_se}
\end{equation}
where $W(\phi)$ is the polar distribution of the Cherenkov sound. It can be
written as the sum,
\begin{equation}
W(\phi)=\sum_i x_i^2(\phi)f_1[x_i(\phi),\phi]f_2[x_i(\phi),\phi],
\label{W_phi}
\end{equation}
over all roots of the energy-momentum conservation equation,
\begin{equation}
\begin{split}
&\sqrt{\biggl(\frac{v}{c}\biggl)^2+\frac{v}{c}h\sin(\phi_\vec{H})+\frac{h^2}{4}}-\\
&-\sqrt{\biggl(\frac{v}{c}\biggl)^2f_3(x,\phi)+\frac{v}{c}hf_4(x,\phi)+\frac{h^2}{4}}-x=0,
\end{split}
\label{En_mom_consv}
\end{equation}
where the physical meaning of $x$ is the ratio $|\vec{q}|/|\vec{p}|$ between
the magnitudes of the phonon and helical particle momenta. In
Eq. (\ref{W_phi}) the functions $f_{1,2}(x,\phi)$ are
\begin{equation}
\begin{split}
&f_1(x,\phi)=\\
&=\frac{1}{2}\left(1+
\frac{\frac{v}{c}(1-x\cos(\phi))+\frac{1}{2}h\sin(\phi_\vec{H})}{\sqrt{\bigl(\frac{v}{c}\bigl)^2
    f_3(x,\phi)+\frac{v}{c}hf_4(x,\phi)+\frac{h^2}{4}}}\right),\\
&f_2(x,\phi)=\\
&=\left[1+\frac{\bigl(\frac{v}{c}\bigl)^2f'_3(x,\phi)+\frac{v}{c}hf'_4(x,\phi)}{2\sqrt{\bigl(\frac{v}{c}\bigl)^2
    f_3(x,\phi)+\frac{v}{c}hf_4(x,\phi)+\frac{h^2}{4}}}\right]^{-1},
\end{split}
\label{f_12}
\end{equation}
where $f'(x,\phi)\equiv\partial_xf(x,\phi)$. In Eqs. (\ref{En_mom_consv}) and
(\ref{f_12}) the quantity $h$ is defined as $h\equiv
g\mu_\text{B}|\vec{H}|/pc$ and the functions $f_{3,4}(x,\phi)$ are
\begin{equation}
\begin{split}
  &f_3(x,\phi)=1+x^2-2x\cos(\phi),\\
  &f_4(x,\phi)=\sin(\phi_\vec{H})+x\sin(-\phi_\vec{H}+\phi).
\end{split}
\label{f_34}
\end{equation}

In Fig. \ref{figure_2} the polar distribution of the Cherenkov sound intensity
$W(\phi)$ on a surface of the three-dimensional topological insulator
Bi$_2$Te$_3$ is shown for the situation depicted in
Fig. \ref{figure_1}. Different values of $h$ correspond to the magnetic field
magnitude such that $|\vec{H}|<H_c$. The critical value
$H_c=2|\vec{p}|v/g\mu_\text{B}$ is obtained from the condition $p_H=|\vec{p}|$
where $p_H=g\mu_\text{B}|\vec{H}|/2v$. The value of $|\vec{p}|=4.05\times
10^{-27}$ kg$\cdot$m/s is chosen to be small enough so that the isotropic
Dirac cone describes well \cite{Fu_2009} the energy-momentum dependence, Eq.
(\ref{Energy_momentum}). Taking the $g$-factor from
Ref. \onlinecite{Rischau_2013}, $g\approx 20$, we obtain for $h=0.0$, $50.0$,
$100.0$, $150.0$ the magnitude of the magnetic field in Tesla,
$|\vec{H}|=0.0$T, $3.1$T, $6.2$T, $9.3$T, respectively. From
Fig. \ref{figure_2} it is clear that the Cherenkov sound is purely forward for
all $|\vec{H}|<H_c$. With the increase of the magnetic field magnitude the
intensity of the Cherenkov sound decreases. At the critical field $H_c=16.9$T
the Cherenkov dissipation fully disappears, {\it i.e.}, $W(\phi)=0$, for all
angles, $-\pi\leqslant\phi<\pi$.

The Cherenkov sound distribution for larger fields, $|\vec{H}|>H_c$, is shown
in Fig. \ref{figure_3}. Here $h=450.0$, $500.0$, $600.0$, $700.0$ which
correspond to $H=27.9$T, $31.0$T, $37.2$T, $43.4$T. As expected from our
qualitative discussion, at $|\vec{H}|>H_c$ the Cherenkov sound is of pure
backward nature. Note that the maximum of the backward sound intensity is at
$\phi=\pm\pi$ which corresponds to strictly backward sound. This means that at
large magnetic fields helical particles mainly scatter strictly forward thus
keeping the tendency to produce backward sound in subsequent scattering unless
they finally reach the Dirac point $p_H$.

For completeness we also present results for $\phi_\vec{H}\neq -\pi/2$. In
Fig. \ref{figure_4} we show the Cherenkov sound distribution for the case
$|\vec{H}|<H_c$ while in Fig. \ref{figure_5} it is shown for the case
$|\vec{H}|>H_c$. In both cases the symmetry of the angular distribution of the
Cherenkov sound breaks and it is no longer of pure forward or backward nature
as it was for $\phi_\vec{H}=-\pi/2$ shown as the black curves in
Figs. \ref{figure_4} and \ref{figure_5}, respectively. In addition to the
forward and backward Cherenkov sound narrow angular droplets of the backward
and forward sound can be seen in Figs. \ref{figure_4} and \ref{figure_5},
respectively.

Finally, we would like to emphasize the asymmetry between the forward and
backward distributions. In particular, as shown in
Ref. \onlinecite{Smirnov_2013}, the strictly forward sound is absent, as can
be seen in Fig. \ref{figure_2}. At the same time the strictly backward sound
in Fig. \ref{figure_3} is maximal. The physical reason for this asymmetry lies
in the helical nature of the particles on a surface of a three-dimensional
topological insulator. More precisely, the spinorial structure of these states
has a strong dependence on the momentum orientation and the sound distribution
depends on the mutual orientation between the momentum $\vec{p}$ and the group
velocity $\vec{v}_g$. The forward and backward sound defined with respect to
the momentum $\vec{p}$ is, of course, always forward with respect to the group
velocity $\vec{v}_g$. However, due to the helical structure of the surface
states, the sound distribution has distinct shapes depending on whether the
vectors $\vec{p}$ and $\vec{v}_g$ are parallel or antiparallel.

This asymmetry suggests an experimental challenge for an implementation of an
electronic device such as an acoustic ratchet. In this device the magnitude of
a periodic in-plane magnetic field slowly changes from zero up to some maximal
value $H_{max}>H_c$ and back to zero. During this period no strictly forward
sound will be produced by helical particles with a fixed momentum whereas they
will produce the strictly backward sound. Thus a directed phonon flow in the
direction opposite to the $x$-axis will be created.

\section{Conclusions}\label{concl}
In conclusion, the Cherenkov sound excited on a surface of a three-dimensional
topological insulator may be effectively controlled by an in-plane magnetic
field. Applying such a field perpendicular to the helical particle propagation
suppresses the Cherenkov dissipation up to zero at a critical field. For
larger fields the Cherenkov sound asymmetrically reverses its direction and
acquires pure backward nature.

The magnetic field shift of the conduction/valence band minimum/maximum in the
momentum space and the dependence of the spinorial structure of the helical
particles on the momentum orientation, are specific to three-dimensional
topological insulators and do not have counterparts in conventional systems
where under the Zeeman split the energy bands shift along the energy axis but
not in the momentum space. As a result, the Cherenkov sound on a surface of a
three-dimensional topological insulator in an in-plane magnetic field is a
unique physical phenomenon totally distinct from what has been known about the
Cherenkov radiation in conventional materials without and with magnetic field
as well as in three-dimensional topological insulators without magnetic
field.

This unique behavior occurs at low energies and, thus, of practical relevance in
the control of the energy dissipation in future low power design electronics
based on topological insulators as well as in building different acoustic devices,
in particular, acoustic ratchets generating directed sound flows.

\section{Acknowledgments}
Support from the DFG under the program SFB 689 is acknowledged.


\begin{thebibliography}{21}%
\makeatletter
\providecommand \@ifxundefined [1]{%
 \@ifx{#1\undefined}
}%
\providecommand \@ifnum [1]{%
 \ifnum #1\expandafter \@firstoftwo
 \else \expandafter \@secondoftwo
 \fi
}%
\providecommand \@ifx [1]{%
 \ifx #1\expandafter \@firstoftwo
 \else \expandafter \@secondoftwo
 \fi
}%
\providecommand \natexlab [1]{#1}%
\providecommand \enquote  [1]{``#1''}%
\providecommand \bibnamefont  [1]{#1}%
\providecommand \bibfnamefont [1]{#1}%
\providecommand \citenamefont [1]{#1}%
\providecommand \href@noop [0]{\@secondoftwo}%
\providecommand \href [0]{\begingroup \@sanitize@url \@href}%
\providecommand \@href[1]{\@@startlink{#1}\@@href}%
\providecommand \@@href[1]{\endgroup#1\@@endlink}%
\providecommand \@sanitize@url [0]{\catcode `\\12\catcode `\$12\catcode
  `\&12\catcode `\#12\catcode `\^12\catcode `\_12\catcode `\%12\relax}%
\providecommand \@@startlink[1]{}%
\providecommand \@@endlink[0]{}%
\providecommand \url  [0]{\begingroup\@sanitize@url \@url }%
\providecommand \@url [1]{\endgroup\@href {#1}{\urlprefix }}%
\providecommand \urlprefix  [0]{URL }%
\providecommand \Eprint [0]{\href }%
\providecommand \doibase [0]{http://dx.doi.org/}%
\providecommand \selectlanguage [0]{\@gobble}%
\providecommand \bibinfo  [0]{\@secondoftwo}%
\providecommand \bibfield  [0]{\@secondoftwo}%
\providecommand \translation [1]{[#1]}%
\providecommand \BibitemOpen [0]{}%
\providecommand \bibitemStop [0]{}%
\providecommand \bibitemNoStop [0]{.\EOS\space}%
\providecommand \EOS [0]{\spacefactor3000\relax}%
\providecommand \BibitemShut  [1]{\csname bibitem#1\endcsname}%
\let\auto@bib@innerbib\@empty
%</preamble>
\bibitem [{\citenamefont {Cherenkov}(1934)}]{Cherenkov}%
  \BibitemOpen
  \bibfield  {author} {\bibinfo {author} {\bibfnamefont {P.~A.}\ \bibnamefont
  {Cherenkov}},\ }\href@noop {} {\bibfield  {journal} {\bibinfo  {journal}
  {Doklady Akad. Nauk SSSR}\ }\textbf {\bibinfo {volume} {2}},\ \bibinfo
  {pages} {451} (\bibinfo {year} {1934})}\BibitemShut {NoStop}%
\bibitem [{\citenamefont {Tamm}\ and\ \citenamefont {Frank}(1937)}]{Tamm}%
  \BibitemOpen
  \bibfield  {author} {\bibinfo {author} {\bibfnamefont {I.~E.}\ \bibnamefont
  {Tamm}}\ and\ \bibinfo {author} {\bibfnamefont {I.~M.}\ \bibnamefont
  {Frank}},\ }\href@noop {} {\bibfield  {journal} {\bibinfo  {journal} {Doklady
  Akad. Nauk. SSSR}\ }\textbf {\bibinfo {volume} {14}},\ \bibinfo {pages} {107}
  (\bibinfo {year} {1937})}\BibitemShut {NoStop}%
\bibitem [{\citenamefont {Komirenko}\ \emph {et~al.}(2000)\citenamefont
  {Komirenko}, \citenamefont {Kim}, \citenamefont {Demidenko}, \citenamefont
  {Kochelap},\ and\ \citenamefont {Stroscio}}]{Komirenko_2000}%
  \BibitemOpen
  \bibfield  {author} {\bibinfo {author} {\bibfnamefont {S.~M.}\ \bibnamefont
  {Komirenko}}, \bibinfo {author} {\bibfnamefont {K.~W.}\ \bibnamefont {Kim}},
  \bibinfo {author} {\bibfnamefont {A.~A.}\ \bibnamefont {Demidenko}}, \bibinfo
  {author} {\bibfnamefont {V.~A.}\ \bibnamefont {Kochelap}}, \ and\ \bibinfo
  {author} {\bibfnamefont {M.~A.}\ \bibnamefont {Stroscio}},\ }\href@noop {}
  {\bibfield  {journal} {\bibinfo  {journal} {Appl.\ Phys.\ Lett.}\ }\textbf
  {\bibinfo {volume} {76}},\ \bibinfo {pages} {1869} (\bibinfo {year}
  {2000})}\BibitemShut {NoStop}%
\bibitem [{\citenamefont {Zhao}\ \emph {et~al.}(2009)\citenamefont {Zhao},
  \citenamefont {Zhang}, \citenamefont {Chen},\ and\ \citenamefont
  {Xu}}]{Zhao_2009}%
  \BibitemOpen
  \bibfield  {author} {\bibinfo {author} {\bibfnamefont {X.~F.}\ \bibnamefont
  {Zhao}}, \bibinfo {author} {\bibfnamefont {J.}~\bibnamefont {Zhang}},
  \bibinfo {author} {\bibfnamefont {S.~M.}\ \bibnamefont {Chen}}, \ and\
  \bibinfo {author} {\bibfnamefont {W.}~\bibnamefont {Xu}},\ }\href@noop {}
  {\bibfield  {journal} {\bibinfo  {journal} {J. Appl. Phys.}\ }\textbf
  {\bibinfo {volume} {105}},\ \bibinfo {pages} {104514} (\bibinfo {year}
  {2009})}\BibitemShut {NoStop}%
\bibitem [{\citenamefont {Willatzen}\ and\ \citenamefont
  {Christensen}(2014)}]{Willatzen_2014}%
  \BibitemOpen
  \bibfield  {author} {\bibinfo {author} {\bibfnamefont {M.}~\bibnamefont
  {Willatzen}}\ and\ \bibinfo {author} {\bibfnamefont {J.}~\bibnamefont
  {Christensen}},\ }\href@noop {} {\bibfield  {journal} {\bibinfo  {journal}
  {Phys.\ Rev.\ B}\ }\textbf {\bibinfo {volume} {89}},\ \bibinfo {pages}
  {041201(R)} (\bibinfo {year} {2014})}\BibitemShut {NoStop}%
\bibitem [{\citenamefont {Zhao}\ \emph {et~al.}(2013)\citenamefont {Zhao},
  \citenamefont {Xu},\ and\ \citenamefont {Peeters}}]{Zhao_2013}%
  \BibitemOpen
  \bibfield  {author} {\bibinfo {author} {\bibfnamefont {C.~X.}\ \bibnamefont
  {Zhao}}, \bibinfo {author} {\bibfnamefont {W.}~\bibnamefont {Xu}}, \ and\
  \bibinfo {author} {\bibfnamefont {F.~M.}\ \bibnamefont {Peeters}},\
  }\href@noop {} {\bibfield  {journal} {\bibinfo  {journal} {Appl.\ Phys.\
  Lett.}\ }\textbf {\bibinfo {volume} {102}},\ \bibinfo {pages} {222101}
  (\bibinfo {year} {2013})}\BibitemShut {NoStop}%
\bibitem [{\citenamefont {Kovrizhin}\ and\ \citenamefont
  {Maksimov}(2001)}]{Kovrizhin_2001}%
  \BibitemOpen
  \bibfield  {author} {\bibinfo {author} {\bibfnamefont {D.~L.}\ \bibnamefont
  {Kovrizhin}}\ and\ \bibinfo {author} {\bibfnamefont {L.~A.}\ \bibnamefont
  {Maksimov}},\ }\href@noop {} {\bibfield  {journal} {\bibinfo  {journal}
  {Phys.\ Lett. A}\ }\textbf {\bibinfo {volume} {282}},\ \bibinfo {pages} {421}
  (\bibinfo {year} {2001})}\BibitemShut {NoStop}%
\bibitem [{\citenamefont {Luo}\ \emph {et~al.}(2003)\citenamefont {Luo},
  \citenamefont {Ibanescu}, \citenamefont {Johnson},\ and\ \citenamefont
  {Joannopoulos}}]{Joannopoulos}%
  \BibitemOpen
  \bibfield  {author} {\bibinfo {author} {\bibfnamefont {C.}~\bibnamefont
  {Luo}}, \bibinfo {author} {\bibfnamefont {M.}~\bibnamefont {Ibanescu}},
  \bibinfo {author} {\bibfnamefont {S.~G.}\ \bibnamefont {Johnson}}, \ and\
  \bibinfo {author} {\bibfnamefont {J.~D.}\ \bibnamefont {Joannopoulos}},\
  }\href@noop {} {\bibfield  {journal} {\bibinfo  {journal} {Science}\ }\textbf
  {\bibinfo {volume} {299}},\ \bibinfo {pages} {368} (\bibinfo {year}
  {2003})}\BibitemShut {NoStop}%
\bibitem [{\citenamefont {Bychkov}\ and\ \citenamefont
  {Rashba}(1984)}]{Bychkov}%
  \BibitemOpen
  \bibfield  {author} {\bibinfo {author} {\bibfnamefont {Y.~A.}\ \bibnamefont
  {Bychkov}}\ and\ \bibinfo {author} {\bibfnamefont {E.~I.}\ \bibnamefont
  {Rashba}},\ }\href@noop {} {\bibfield  {journal} {\bibinfo  {journal} {JETP
  Letters}\ }\textbf {\bibinfo {volume} {39}},\ \bibinfo {pages} {78} (\bibinfo
  {year} {1984})}\BibitemShut {NoStop}%
\bibitem [{\citenamefont {Smirnov}(2011)}]{Smirnov_2011}%
  \BibitemOpen
  \bibfield  {author} {\bibinfo {author} {\bibfnamefont {S.}~\bibnamefont
  {Smirnov}},\ }\href@noop {} {\bibfield  {journal} {\bibinfo  {journal}
  {Phys.\ Rev.\ B}\ }\textbf {\bibinfo {volume} {83}},\ \bibinfo {pages}
  {081308(R)} (\bibinfo {year} {2011})}\BibitemShut {NoStop}%
\bibitem [{\citenamefont {Hasan}\ and\ \citenamefont
  {Kane}(2010)}]{Hasan_2010}%
  \BibitemOpen
  \bibfield  {author} {\bibinfo {author} {\bibfnamefont {M.~Z.}\ \bibnamefont
  {Hasan}}\ and\ \bibinfo {author} {\bibfnamefont {C.~L.}\ \bibnamefont
  {Kane}},\ }\href@noop {} {\bibfield  {journal} {\bibinfo  {journal} {Rev.\
  Mod.\ Phys.}\ }\textbf {\bibinfo {volume} {82}},\ \bibinfo {pages} {3045}
  (\bibinfo {year} {2010})}\BibitemShut {NoStop}%
\bibitem [{\citenamefont {Qi}\ and\ \citenamefont {Zhang}(2011)}]{Qi_2011}%
  \BibitemOpen
  \bibfield  {author} {\bibinfo {author} {\bibfnamefont {X.-L.}\ \bibnamefont
  {Qi}}\ and\ \bibinfo {author} {\bibfnamefont {S.-C.}\ \bibnamefont {Zhang}},\
  }\href@noop {} {\bibfield  {journal} {\bibinfo  {journal} {Rev.\ Mod.\
  Phys.}\ }\textbf {\bibinfo {volume} {83}},\ \bibinfo {pages} {1057} (\bibinfo
  {year} {2011})}\BibitemShut {NoStop}%
\bibitem [{\citenamefont {Wu}\ \emph {et~al.}(2006)\citenamefont {Wu},
  \citenamefont {Bernevig},\ and\ \citenamefont {Zhang}}]{Wu_2006}%
  \BibitemOpen
  \bibfield  {author} {\bibinfo {author} {\bibfnamefont {C.}~\bibnamefont
  {Wu}}, \bibinfo {author} {\bibfnamefont {B.~A.}\ \bibnamefont {Bernevig}}, \
  and\ \bibinfo {author} {\bibfnamefont {S.-C.}\ \bibnamefont {Zhang}},\
  }\href@noop {} {\bibfield  {journal} {\bibinfo  {journal} {Phys.\ Rev.\
  Lett.}\ }\textbf {\bibinfo {volume} {96}},\ \bibinfo {pages} {106401}
  (\bibinfo {year} {2006})}\BibitemShut {NoStop}%
\bibitem [{\citenamefont {Xu}\ and\ \citenamefont {Moore}(2006)}]{Xu_2006}%
  \BibitemOpen
  \bibfield  {author} {\bibinfo {author} {\bibfnamefont {C.}~\bibnamefont
  {Xu}}\ and\ \bibinfo {author} {\bibfnamefont {J.~E.}\ \bibnamefont {Moore}},\
  }\href@noop {} {\bibfield  {journal} {\bibinfo  {journal} {Phys.\ Rev.\ B}\
  }\textbf {\bibinfo {volume} {73}},\ \bibinfo {pages} {045322} (\bibinfo
  {year} {2006})}\BibitemShut {NoStop}%
\bibitem [{\citenamefont {Smirnov}(2013)}]{Smirnov_2013}%
  \BibitemOpen
  \bibfield  {author} {\bibinfo {author} {\bibfnamefont {S.}~\bibnamefont
  {Smirnov}},\ }\href@noop {} {\bibfield  {journal} {\bibinfo  {journal}
  {Phys.\ Rev.\ B}\ }\textbf {\bibinfo {volume} {88}},\ \bibinfo {pages}
  {205301} (\bibinfo {year} {2013})}\BibitemShut {NoStop}%
\bibitem [{\citenamefont {Fu}(2009)}]{Fu_2009}%
  \BibitemOpen
  \bibfield  {author} {\bibinfo {author} {\bibfnamefont {L.}~\bibnamefont
  {Fu}},\ }\href@noop {} {\bibfield  {journal} {\bibinfo  {journal} {Phys.\
  Rev.\ Lett.}\ }\textbf {\bibinfo {volume} {103}},\ \bibinfo {pages} {266801}
  (\bibinfo {year} {2009})}\BibitemShut {NoStop}%
\bibitem [{\citenamefont {Smirnov}\ \emph {et~al.}(2008)\citenamefont
  {Smirnov}, \citenamefont {Bercioux}, \citenamefont {Grifoni},\ and\
  \citenamefont {Richter}}]{Smirnov_2008}%
  \BibitemOpen
  \bibfield  {author} {\bibinfo {author} {\bibfnamefont {S.}~\bibnamefont
  {Smirnov}}, \bibinfo {author} {\bibfnamefont {D.}~\bibnamefont {Bercioux}},
  \bibinfo {author} {\bibfnamefont {M.}~\bibnamefont {Grifoni}}, \ and\
  \bibinfo {author} {\bibfnamefont {K.}~\bibnamefont {Richter}},\ }\href@noop
  {} {\bibfield  {journal} {\bibinfo  {journal} {Phys.\ Rev.\ Lett.}\ }\textbf
  {\bibinfo {volume} {100}},\ \bibinfo {pages} {230601} (\bibinfo {year}
  {2008})}\BibitemShut {NoStop}%
\bibitem [{\citenamefont {{Castro Neto}}\ \emph {et~al.}(2009)\citenamefont
  {{Castro Neto}}, \citenamefont {Guinea}, \citenamefont {Peres}, \citenamefont
  {Novoselov},\ and\ \citenamefont {Geim}}]{Castro_Neto_2009}%
  \BibitemOpen
  \bibfield  {author} {\bibinfo {author} {\bibfnamefont {A.~H.}\ \bibnamefont
  {{Castro Neto}}}, \bibinfo {author} {\bibfnamefont {F.}~\bibnamefont
  {Guinea}}, \bibinfo {author} {\bibfnamefont {N.~M.~R.}\ \bibnamefont
  {Peres}}, \bibinfo {author} {\bibfnamefont {K.~S.}\ \bibnamefont
  {Novoselov}}, \ and\ \bibinfo {author} {\bibfnamefont {A.~K.}\ \bibnamefont
  {Geim}},\ }\href@noop {} {\bibfield  {journal} {\bibinfo  {journal} {Rev.\
  Mod.\ Phys.}\ }\textbf {\bibinfo {volume} {81}},\ \bibinfo {pages} {109}
  (\bibinfo {year} {2009})}\BibitemShut {NoStop}%
\bibitem [{\citenamefont {Abrikosov}\ \emph {et~al.}(1963)\citenamefont
  {Abrikosov}, \citenamefont {Gorkov},\ and\ \citenamefont
  {Dzyaloshinski}}]{AGD}%
  \BibitemOpen
  \bibfield  {author} {\bibinfo {author} {\bibfnamefont {A.~A.}\ \bibnamefont
  {Abrikosov}}, \bibinfo {author} {\bibfnamefont {L.~P.}\ \bibnamefont
  {Gorkov}}, \ and\ \bibinfo {author} {\bibfnamefont {I.~E.}\ \bibnamefont
  {Dzyaloshinski}},\ }\href@noop {} {\emph {\bibinfo {title} {Methods of
  Quantum Field Theory in Statistical Physics}}}\ (\bibinfo  {publisher}
  {Dover, New York},\ \bibinfo {year} {1963})\BibitemShut {NoStop}%
\bibitem [{\citenamefont {Levitov}\ and\ \citenamefont {Shitov}(2003)}]{LS}%
  \BibitemOpen
  \bibfield  {author} {\bibinfo {author} {\bibfnamefont {L.~S.}\ \bibnamefont
  {Levitov}}\ and\ \bibinfo {author} {\bibfnamefont {A.~V.}\ \bibnamefont
  {Shitov}},\ }\href@noop {} {\emph {\bibinfo {title} {Green's Functions.
  Problems and Solutions}}},\ \bibinfo {edition} {2nd}\ ed.\ (\bibinfo
  {publisher} {Fizmatlit, Moscow},\ \bibinfo {year} {2003})\ \bibinfo {note}
  {in Russian}\BibitemShut {NoStop}%
\bibitem [{\citenamefont {Rischau}\ \emph {et~al.}(2013)\citenamefont
  {Rischau}, \citenamefont {Leridon}, \citenamefont {Fauqu{\'e}}, \citenamefont
  {Metayer},\ and\ \citenamefont {{van der Beek}}}]{Rischau_2013}%
  \BibitemOpen
  \bibfield  {author} {\bibinfo {author} {\bibfnamefont {C.~W.}\ \bibnamefont
  {Rischau}}, \bibinfo {author} {\bibfnamefont {B.}~\bibnamefont {Leridon}},
  \bibinfo {author} {\bibfnamefont {B.}~\bibnamefont {Fauqu{\'e}}}, \bibinfo
  {author} {\bibfnamefont {V.}~\bibnamefont {Metayer}}, \ and\ \bibinfo
  {author} {\bibfnamefont {C.~J.}\ \bibnamefont {{van der Beek}}},\ }\href@noop
  {} {\bibfield  {journal} {\bibinfo  {journal} {Phys.\ Rev.\ B}\ }\textbf
  {\bibinfo {volume} {88}},\ \bibinfo {pages} {205207} (\bibinfo {year}
  {2013})}\BibitemShut {NoStop}%
\end{thebibliography}
\end{document}